\documentstyle{mn}

\newif\ifAMStwofonts

\ifoldfss
  \ifCUPmtlplainloaded \else
    \NewTextAlphabet{textbfit} {cmbxti10} {}
    \NewTextAlphabet{textbfss} {cmssbx10} {}
    \NewMathAlphabet{mathbfit} {cmbxti10} {} 
    \NewMathAlphabet{mathbfss} {cmssbx10} {} 
  \fi
  \ifAMStwofonts
    \ifCUPmtlplainloaded \else
      \NewSymbolFont{upmath} {eurm10}
      \NewSymbolFont{AMSa} {msam10}
      \NewMathSymbol{\upi}     {0}{upmath}{19}
      \NewMathSymbol{\umu}     {0}{upmath}{16}
      \NewMathSymbol{\upartial}{0}{upmath}{40}
      \NewMathSymbol{\leqslant}{3}{AMSa}{36}
      \NewMathSymbol{\geqslant}{3}{AMSa}{3E}

    \fi
  \fi
\fi 

\ifnfssone
  \newmathalphabet{\mathit}
  \addtoversion{normal}{\mathit}{cmr}{m}{it}
  \addtoversion{bold}{\mathit}{cmr}{bx}{it}
  \newmathalphabet{\mathbfit} 
  \addtoversion{normal}{\mathbfit}{cmr}{bx}{it}
  \addtoversion{bold}{\mathbfit}{cmr}{bx}{it}
  \newmathalphabet{\mathbfss} 
  \addtoversion{normal}{\mathbfss}{cmss}{bx}{n}
  \addtoversion{bold}{\mathbfss}{cmss}{bx}{n}
  \ifAMStwofonts
    \ifCUPmtlplainloaded \else
      \UseAMStwoboldmath
      \makeatletter
      \new@mathgroup\upmath@group
      \define@mathgroup\mv@normal\upmath@group{eur}{m}{n}
      \define@mathgroup\mv@bold\upmath@group{eur}{b}{n}
      \edef\UPM{\hexnumber\upmath@group}
      \new@mathgroup\amsa@group
      \define@mathgroup\mv@normal\amsa@group{msa}{m}{n}
      \define@mathgroup\mv@bold\amsa@group{msa}{m}{n}
      \edef\AMSa{\hexnumber\amsa@group}
      \makeatother
      \mathchardef\upi="0\UPM19
      \mathchardef\umu="0\UPM16
      \mathchardef\upartial="0\UPM40
      \mathchardef\leqslant="3\AMSa36
      \mathchardef\geqslant="3\AMSa3E
    \fi
  \fi
\fi 

\ifnfsstwo
  \DeclareMathAlphabet{\mathbfit}{OT1}{cmr}{bx}{it}
  \SetMathAlphabet\mathbfit{bold}{OT1}{cmr}{bx}{it}
  \DeclareMathAlphabet{\mathbfss}{OT1}{cmss}{bx}{n}
  \SetMathAlphabet\mathbfss{bold}{OT1}{cmss}{bx}{n}
  \ifAMStwofonts
    \ifCUPmtlplainloaded \else
      \DeclareSymbolFont{UPM}{U}{eur}{m}{n}
      \SetSymbolFont{UPM}{bold}{U}{eur}{b}{n}
      \DeclareSymbolFont{AMSa}{U}{msa}{m}{n}
      \DeclareMathSymbol{\upi}{0}{UPM}{"19}
      \DeclareMathSymbol{\umu}{0}{UPM}{"16}
      \DeclareMathSymbol{\upartial}{0}{UPM}{"40}
      \DeclareMathSymbol{\leqslant}{3}{AMSa}{"36}
      \DeclareMathSymbol{\geqslant}{3}{AMSa}{"3E}
    \fi
  \fi
\fi 

\ifCUPmtlplainloaded \else
  \ifAMStwofonts \else 
    \def\upi{\pi}
    \def\umu{\mu}
    \def\upartial{\partial}
  \fi
\fi

\title{Orbital solution for the MACHO$^\ast$05:34:41.3\,-69:31:39 
O3\,If$^{\ast}+$O6:V eclipsing binary system in the LMC}
\author[P.G. Ostrov]
{P.G. Ostrov \\ Facultad de Ciencias Astron\'omicas y Geof\'{\i}sicas, 
Universidad Nacional de La Plata, Paseo del Bosque S/N, 1900 La Plata, 
Argentina}

\date{Accepted 1988 December 15.
      Received 1988 December 14;
      in original form 1988 October 11}

\pagerange{\pageref{firstpage}--\pageref{lastpage}}
\pubyear{1994}

\newcommand{\pepita}{MACHO$^\ast$05:34:41.3\,-69:31:39 }

\begin{document}

\maketitle

\begin{abstract}
An orbital solution for the \pepita
~eclipsing binary system is presented, based on the published light curve and 
spectral data obtained with the 2.15-m telescope at CASLEO.
Based on these spectroscopic observations, this system's binary
components were classified as O3\,If$^{\ast}$ and O6:V respectively. 
The radial velocity data along with the published light curve were analysed
with the Wilson-Devinney code to derive  the following masses and radii 
for the components of this system:
$M_1=41 \pm 1.2 \, {\rm M}_{\sun}$, $R_1=9.6 \pm
0.02 \, {\rm R}_{\sun}$, $M_2=27 \pm 1.2 \, {\rm M}_{\sun}$ and $R_2=8.0 
\pm 0.05 \, {\rm R}_{\sun}$.
The solution shows that the system is in overcontact as one would expect from 
the derived masses and the very short orbital period ($\sim$ 1.4 days).
\end{abstract}


\maketitle  

\section{Introduction}

One of the consequences of the huge amount of data produced by
the MACHO project was the discovery of more than six hundred new eclipsing
binary stars in the Large Magellanic Cloud (Alcock et al. 1997).  
Given  that  the  brightest
among these systems lay within the scope of moderate  size  telescopes
for intermediate dispersion spectroscopy, we began to  acquire  spectra  of 
those objects using the $2.15-m$ telescope at  CASLEO\footnote{
Complejo Astron\'omico El Leoncito, operated under agreement between 
the Consejo Nacional de Investigaciones Cient\'{\i}ficas y T\'ecnicas de la 
Rep\'ublica Argentina and the National Universities of La Plata, C\'ordoba and 
San Juan.} (San Juan, Argentina), with the aim of producing determinations of 
physical parameters for them.
One  of  the observed stars, the so-called \pepita, 
located in the 
H{\sc ii} region DEM\,242 (Davies, Elliott \& Meaburn 1976) turned out to be  
a double lined O3\,If$^{\ast}$+O6:V system, a spectral type we 
would not expect for such a short period binary. Apart from the general surveys
as the Digitized Sky Survey, there are no previous observations of this star.

There are an intrinsically small number of early O-type stars, and 
an even smaller number of them  in eclipsing binary systems. 
No O3 star is known to be an eclipsing binary to date.
Besides,  the masses predicted from comparison of the luminosities
of these very hot stars with  numerical models are of the order of 
$100\,{\rm M}_{\sun}$, a value much larger than that inferred 
from observations of binary stars of similar spectral types.
For these reasons, each newly discovered O3 star presents an  opportunity  to
provide new data that contributes to resolving this problem.

Of course, it must be considered that numerical models deal 
with single and isolated stars, while the binary studied here is an 
extremely  close system,  with  a  period of $\sim 1.4$ days and undoubtedly
subject to very strong interaction effects.
Consequently, the masses derived here must be considered with caution
when compared with evolutionary models.

\section{Observations and data analysis}
  
\subsection{Photometric and spectroscopic data}

The photometric light curve for \pepita
was published by Alcock et al. (1997,1997a).  
Data are available for two bands, namely $V$ and $R$. The $V$ magnitude at
maximum is 13.54, while primary and secondary minima have $V=13.91$ and 13.89
respectively. The $V-R$ color of the system is $-0.15$ (in the Kron-Cousins 
photometric system), with no noticeable changes with phase. Uncertainties 
given by Alcock et al. rise to 0.07 in $V$ and 0.03 in $V-R$

During an observing run in January 2000 with  the $2.15-m$  telescope  at
CASLEO (San Juan, Argentina), N.I. Morrell acquired 8 spectra near the times 
of quadratures.

\begin{figure}

 \vspace{6cm}

 \begin{picture}(0,0) 
  \put(-72,-252){\includegraphics{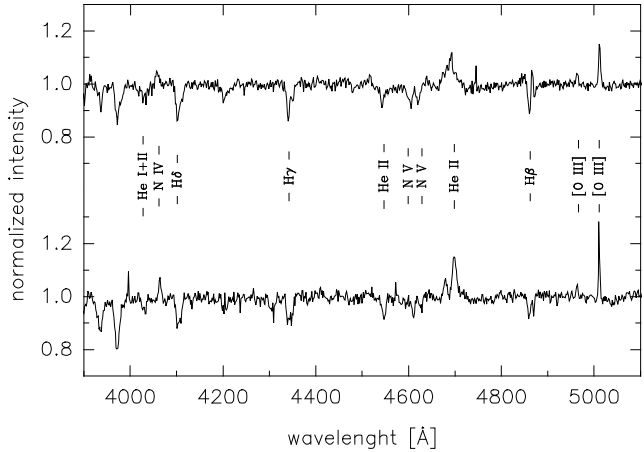}} 
 \end{picture}

 \caption{CCD spectrograms of MACHO$^\ast$05:34:41.3\,-69:31:39
  obtained at CASLEO, corresponding to 
  both quadratures.}
\end{figure}

A REOSC spectrograph  was used in its single dispersion mode, with a 600 
grooves mm$^{-1}$ grating and a Tek $1024\times 1024$ CCD as detector, 
this instrumental configuration giving a reciprocal dispersion 
of 1.8 \AA\, pixel$^{-1}$ in the blue-yellow region of the spectrum.
The spectrograms covered the region from 3800\,\AA ~to 5440\,\AA, with a 
typical resolution (as measured from the  He-Ar  comparison  lamp lines) of 2
pixels and the signal to noise ratio ranged from 50 to 100.

The usual sets of bias, dark and flat-field  frames were   
acquired during each observing night. The spectroscopic observations were 
processed  and  analysed with  IRAF \footnote {IRAF software is distributed by 
NOAO, operated by  AURA  for NSF} routines.
In order to check the system's stability for radial velocity measurements,
radial velocities for the nebular [O{\sc iii}] emission lines
in our spectra were derived, obtaining an average of +258 km\,s$^{-1}$ with a
standard deviation of 22 km\,s$^{-1}$, a radial velocity that is well
within the expected range for LMC objects. 

Just a quick look to the obtained spectrograms revealed that the primary star
is an O3\,If$^{\ast}$ presenting strong N{\sc v} 4603 - 4620\,\AA~ absorption 
lines and  emissions of N{\sc iv}, He {\sc ii} and very weak N{\sc iii}. The  
spectrum of the secondary is hard to classify due to relatively low 
dispersion data, but considering those observations where the lines are better 
resolved, it was classified as O6:V. Both components were classified following
the criteria described in Walborn \& Fitzpatrick (1990).
Spectra of \pepita obtained at both quadratures are 
shown in Fig. 1.

To obtain the radial velocities, the N {\sc v} $\lambda$4603 - 4620 \AA\ 
absorptions and the  N{\sc iv} 4058 \AA\ emission were considered as 
representative of the motion of the primary O3\,If$^{\ast}$ component,
while for the radial velocities of the secondary component several He{\sc i} 
absorptions were averaged. The measurements were performed through 
single Gaussian fitting with IRAF routines.

%
\begin{table}
  \caption{Observed  radial velocities of the eclipsing binary \pepita}
  \begin{tabular}{@{}rrrr@{}}
HJD & phase \footnote{Phases were calculated with the ephemerides given by 
Alcock et al. (1997).} & primary & secondary\cr
$2.5 \times 10^6 +$ & & $V_R [km s^{-1}]$ & $V_R [km s^{-1}]$\cr

1555.604 & 0.799 &    521  &  -127  \cr
1557.588 & 0.211 &    -7   &  659   \cr
1562.745 & 0.882 &    447  &  -28   \cr
1563.757 & 0.603 &    428  &  ---   \cr
1564.613 & 0.212 &    -47  &  ---   \cr
1566.712 & 0.706 &    495  &  -165  \cr
1567.622 & 0.354 &    55   &  633   \cr
1571.600 & 0.185 &   -24   &  613   \cr

\end{tabular}
\end{table}

\subsection{Light and radial velocity curve solution}

\begin{figure}
 \vspace{6cm}

 \begin{picture}(0,0) 
  \put(-64,-282){\includegraphics{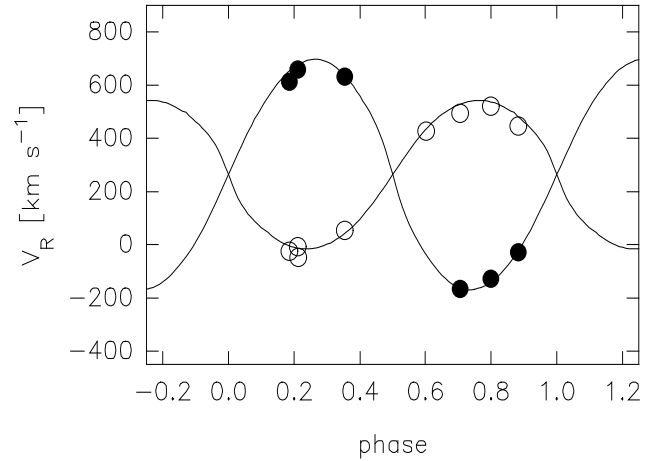}} 
 \end{picture}

 \caption{Observed and modelled radial velocity curve for \pepita. 
          Open circles correspond to the primary 
  component and filled ones stand for the secondary.}
\end{figure}

The radial velocities derived for the quadratures (Table 1) together with the 
published photometric light curve were analysed by means of the 
Wilson-Devinney code (Wilson \& Devinney 1971, Wilson 1990), that 
is very suitable to model binaries with  Roche-lobe filling stars. 

From Schmidt-Kaler (1982), a temperature of 50000 K was assigned for the 
O3\,If$^{\ast}$ primary component. The bolometric albedos where set to A=1.0 
(Rucinski 1969), and gravity brightening coefficients of value g=1.0 
(Lucy 1976) were used. These quantities are the usual guesses for 
radiative envelopes. Linear limb-darkening coefficients were obtained from 
van Hamme (1993). A circular orbit was assumed, according to that arise from
the examination of the light curves and is expected for such a close binary.
All these parameters, together with the ephemerides
supplied by Alcock et al. (1997) ($P=1.404\,740, 
~E_0=2\,449\,073.7109$), were kept constant. 

First, the semimajor axis $a$, the systemic radial velocity 
$V_{\gamma}$ and the mass-ratio $q$ were adjusted using only the radial 
velocity measurements. Thereupon, these parameters were fixed and 
the orbital inclination $i$, the temperature of the secondary star $T_2$, the 
luminosity of the primary $L_1$ and both potentials $\Omega_1$ and $\Omega_2$ 
were fitted using the light curves. Then these two steps were repeated 
iteratively until the solution converged. 
Given that after a few iterations with the differential 
corrections program (DC), the solution evolved to an overcontact configuration,
the program operation was set to mode 1. In this mode, the surface 
potentials of both stars are kept identical ($\Omega_1=\Omega_2$), the same
as the gravity brightening coefficients ($g_1=g_2$), the bolometric albedos
($A_1=A_2$) and the limb-darkening coefficients ($x_1=x_2$). Besides, the polar
temperature of the secondary is set by the gravity brightening law of the 
entire common envelope. Consequently, in the successive iterations only the
parameters $i$, $L_1$ and $\Omega_1$ (together with $a$, $V_{\gamma}$ and $q$)
were fitted. The iterations were stopped when the corrections supplied by the
DC program were smaller than their own errors. Fig. 2 and Fig. 3 show the 
computed $V$ light and radial velocity curves, 
together with the observational data. 

The adopted and fitted model parameters and the resulting stellar dimensions 
are listed in Tables 2 and 3, respectively. Fig. 4 displays the aspect of the 
system at different phases.

\begin{figure}
 \vspace{6cm}

 \begin{picture}(0,0) 
  \put(-18,-220){\includegraphics{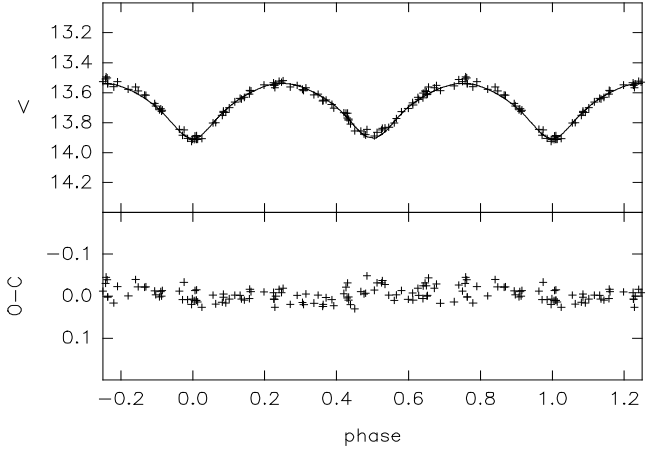}} 
 \end{picture}

 \caption{Top: Observed and modelled $V$ light curve for \pepita. 
         Bottom: (O-C) residuals for the light 
 curve.}
\end{figure}

\begin{table}
  \caption{Model Parameters}
  \begin{tabular}{@{}ll@{}}
 
$T_1$                  &$ 50000\,{\rm K}$ ~(adopted) \cr  
$g_1, g_2$                  & 1.00 ~(adopted)   \cr  
$A_1, A_2$                  & 1.00 ~(adopted)   \cr  
$x_1, x_2$               & 0.616~ (adopted)  \cr
$e$			& 0.00~ (adopted) \cr
\noalign{\medskip} 
$a$                    &$ 22.2 \pm 0.22 \, {\rm R}_{\sun}  $ \cr  
$V_{\gamma}$           &$ 263.1 \pm 3 \, {\rm km s}^{-1}    $ \cr  
$q~(M_2/M_1)$          &$ 0.64 \pm 0.01    $ \cr 
$i$                    &$ 67 \pm 1 \degr      $ \cr  
$\Omega_1, \Omega_2$   &$ 3.046 \pm 0.03   $ \cr
$T_2$                  &$ 49490 \, {\rm K}$ \cr   
\end{tabular}
\end{table}

%
\begin{table}
  \caption{Star Dimensions}
  \begin{tabular}{@{}ll@{}}
 
$M_1       $&$  41.16 \pm 1.2~{\rm M}_{\sun}  $\cr   
$R_1       $&$  9.56 \pm 0.02~{\rm R}_{\sun}  $\cr  
$M_{bol~1} $&$ -9.49     $\cr        
$\log g_1 [cgs]  $&$ 4.09 \pm 0.02        $\cr  
\noalign{\smallskip}  
$M_2   $&$  27.04 \pm 1.2~ {\rm M}_{\sun}  $\cr   
$ R_2  $&$  7.99 \pm 0.05 {\rm R}_{\sun}   $\cr  
$ M_{bol~2} $&$ -9.05   $\cr        
$\log g_2 [cgs]   $&$ 4.07 \pm 0.02   $\cr  
\end{tabular}
\end{table}


\begin{figure}
 \vspace{17cm}

 \begin{picture}(0,0) 
  \put(-70,-58){\includegraphics{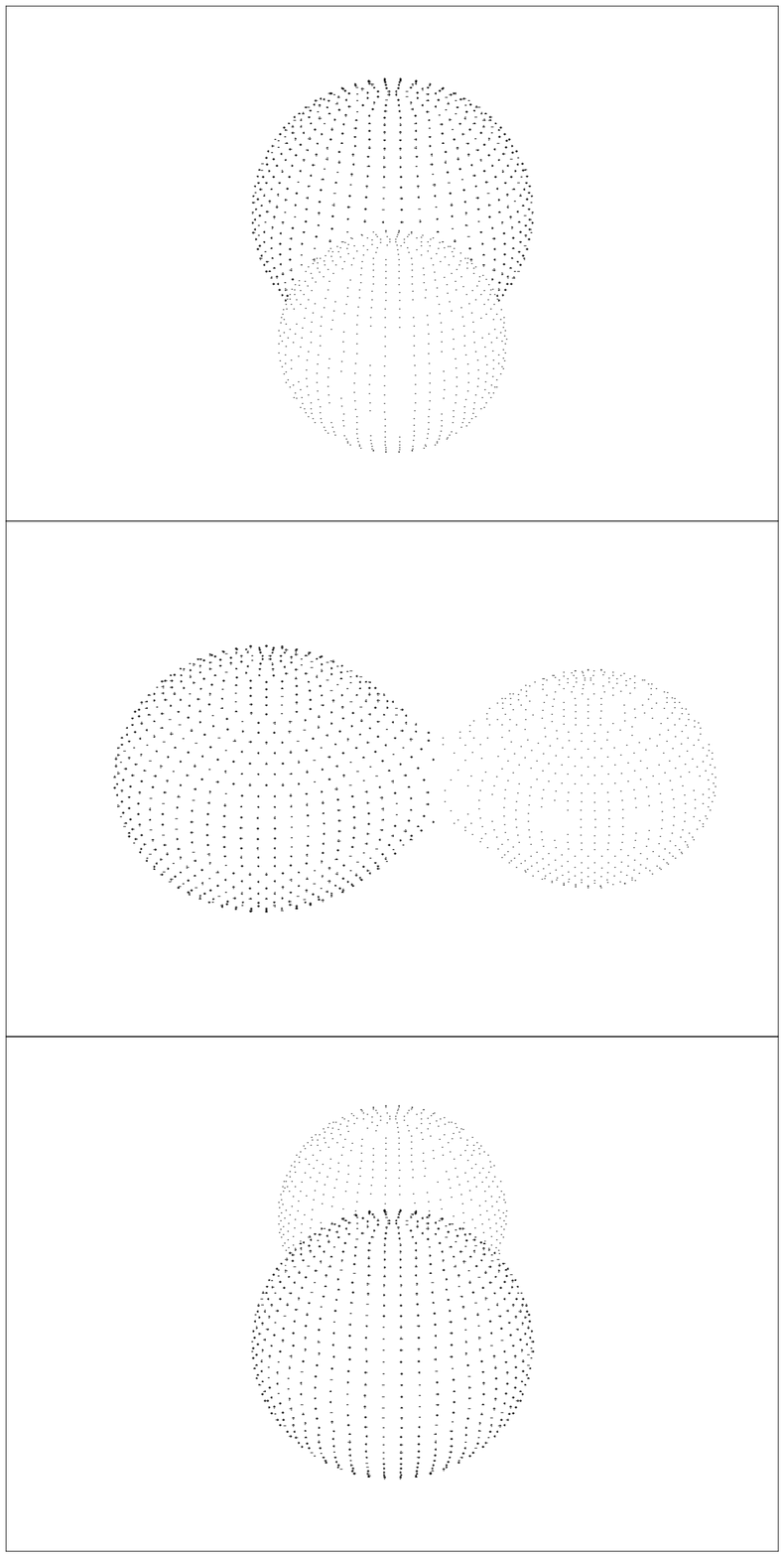}} 
 \end{picture}

 \caption{The aspect of the \pepita 
          overcontact system. Top: at primary minimum, middle: 
          at first quadrature and bottom: at secondary minimum.}
\end{figure}

\subsection{Error Estimates}

The estimation of the parameters' uncertainties was attempted in two ways: 
First, a last run allowing the DC program to fit all the free parameters
(i.e., $a$, $V_{\gamma}$, $i$, $\Omega_1$, $q$ and $L_1$) at one time was
performed, and then the standard deviations of the differential corrections 
supplied by the DC program were used as error estimates. 
Second, the differences between two independent fits, performed 
using the radial velocity data together with only the $V$ or $R$ light curve
separately, was used as error estimate. These differences were
smaller than the errors estimated by the first method, which are quoted 
in tables 2 and 3.

Weighing the uncertainties of the photometric and spectroscopic
data, these error estimates seem rather low. 
This is not surprising, since in operation mode 1 many parameter constraints 
are applied, leaving only the six above quoted quantities to be adjusted.
Taking into account this fact, our resulting parameters (and mainly our error
estimates) must be considered with caution, since  this very  close
system, with a $\sim 1.4$ day orbital period and an O3 star, surely presents
appreciable departures from hydrostatic equilibrium geometry due to  the  
strong stellar winds and radiation pressure.  

Since this star probably belongs to an obscured tight star cluster (the 
resolution of the Digitized Sky Survey and the MACHO plates does not allow
to address this matter), also it was attempted to include a third light in
modelling the light curves. For the best fit, the third light represents only
$\sim 2.6 \%$ of the flux at quadratures, and the other parameters of the model
do not change meaningly.

Looking at the (O-C) residuals in Fig. 2, two slight systematic trends can 
be observed. On one hand, the secondary minimum seems not as deep as it is 
modelled by the Wilson-Devinney program. This feature indicates a bigger 
temperature difference between the components, in accord with the 
spectral types.
The mode of operation 3 of the DC program retains all the parameter
constraints for overcontact systems, except for the secondary's 
temperature $T_2$. An attempt to estimate the secondary temperature from the 
light curves yielded $T_2=47200 \pm 1700$\,K (for $T_1=50000$\,K). However, 
since the difference in surface temperatures affects
mainly the far UV brightness in such very hot stars (in which it is urgent
to acquire data) and its influence in the $V$ and $R$ bands is slight, 
this temperature estimation may be considered with caution. Taken into account
the photometric errors, a temperature difference greater than 5000\,K between
the two components can be rejected.
The other systematic 
deviation of our modelled light curve (very slight also) is an excess of 
brightness in the maximum following the secondary minimum (O'Connell effect, 
see Davidge \& Milone 1984). The O'Connell effect is often modelled
adding ``hot spots'' on the surface of the stars, although its physical causes
in very hot binaries are not clear.
It was not attempted to be modelled this feature, since it is also 
hardly noticeable in comparison with the photometric errors.

\section{Results}

It was an absolutely unexpected discovery to find an O3\,If$^\ast$-type
star in such a very close binary pair. There are not published evolutionary 
models suitable for this system, and single-star models are not adequate.

The derived masses ($\sim 41$ and $\sim 27\,{\rm M}_{\sun} $, with an
inclination of $67 \degr$) seem to be surprisingly low. 

Alcock et al. (1997) presented a fit of the light curves performed by
means of the Nelson-Davis-Etzel model (Nelson \& Davis 1972, Popper \&
Etzel 1981). Assuming a mass-ratio $q=0.95$, they derived an 
inclination of $\sim 57\degr$. 
The difference with the results presented here is not surprising,
given that the Nelson-Davis-Etzel model is not adequate for highly distorted 
systems and the value of $q$ derived from the radial velocity 
measurements is significantly different from the one assumed in Alcock et al.


The system resembles the O4f+O6V binary Sk-67$\degr$105
(see Niemela \& Morrell 1986, Haefner et al. 1994) but it is 
hotter and closer. On the other hand, our mass estimate for the primary 
is similar to that derived by Antokhina et al. (2000) for the O3 star
of the non-eclipsing system  HD\,93205 ($\sim 45$\, M$_{\sun}$ for their best 
fit). 

Undoubtedly, this system will require further investigation, 
especially UV photometry and high dispersion spectroscopy.

\section*{Acknowledgments}

It is a pleasure to acknowledge Dr. Nidia Morrell for observing this object
and providing me with her results. 
This research has made use of the Astronomical Data Center 
catalogs. 


\begin{thebibliography}{}
  
\bibitem{b} Alcock et al., 1997, AJ 114, 326
\bibitem{b} Alcock et al., 1997a, AAS CD-ROM Series Vol. 9
\bibitem{b} Antokhina E.A., Moffat A.F.J., Antokhin I.I., 
Bertrand J-F, Lamontagne, R., 2000, ApJ 529, 463
\bibitem{b} Davidge T.J., Milone E.F., 1984, ApJS 55, 571
\bibitem{b} Davies R.D., Elliot K.H., Meaburn J., 1976, Mem. R. Astron.
Soc. 81, 89
\bibitem{b} Haefner R., Simon K.P., Fiedler A., 1994, A\&A 288, L9
\bibitem{b} van Hamme W., 1993, AJ 106, 2096
\bibitem{b} Lucy L.B., 1976, ApJ 205, 208
\bibitem{b} Nelson, B., Davis W.D., 1972, ApJ 174, 617 
\bibitem{b} Niemela V.S., Morrell N.I., 1986, ApJ 310, 715
\bibitem{b} Niemela V.S., Morrell N.I., Ostrov, P.G, in preparation.
\bibitem{b} Popper, D.M., Etzel P.B., 1981, AJ 86, 102
\bibitem{b} Schmidt-Kaler Th., 1982, in Landolt-B\"ornstein, 
New Series, ed. K. Shaifers \& H.H. Voigt, Group VI, Vol. 2/b
\bibitem{b} Rucinski S.M., 1969, Acta Astron. 19, 245  
\bibitem{b} Walborn N.R., Fitzpatrick E.L., 1990, PASP 379, 411 
\bibitem{b} Wilson R.E., 1990, ApJ 356, 613  
\bibitem{b} Wilson R.E., Devinney E.J., 1971, ApJ 166, 605  
\end{thebibliography}
\end{document}